\documentclass[authoryear,preprint,12pt]{elsarticle}

\usepackage{amssymb}

\usepackage{amsmath}
\usepackage{subfigure}
\usepackage{bm}
\usepackage[hyphens]{url}
\usepackage{color}
\usepackage{here}

\newcommand{\redA}[1]{\textcolor{black}{#1}}
\newcommand{\blueA}[1]{\textcolor{black}{#1}}

\usepackage{setspace}
\journal{Urban Climate}

\begin{document}

\begin{frontmatter}



\title{Two-Stage Super-Resolution Simulation Method of Three-Dimensional Street-Scale Atmospheric Flows for Real-Time Urban Micrometeorology Prediction}


\affiliation[inst1]{
    organization={\redA{Supercomputing Research Center, Institute of Integrated Research, Institute of Science Tokyo}},
    addressline={2-12-1 Ookayama, Meguro-ku},
    city={Tokyo},
    postcode={1528550},
    country={Japan}
}

\author[inst1]{Yuki Yasuda\fnref{corresponding_authoer_info}}
\ead{yasuda.yuki@scrc.iir.isct.ac.jp}
\fntext[Corresponding_authoer_info]{Corresponding author. 2-12-1-i9-111 Ookayama, Meguro-ku, Tokyo, 1528550, Japan. Tel.: +81 3 57343173 (office). Email address: \redA{yasuda.yuki@scrc.iir.isct.ac.jp}}

\author[inst1]{Ryo Onishi}
\ead{onishi.ryo@scrc.iir.isct.ac.jp}

\begin{abstract}
A two-stage super-resolution simulation method is proposed for street-scale air temperature and wind velocity, which considerably reduces computation time while maintaining accuracy. The first stage employs a convolutional neural network (CNN) to correct large-scale flows above buildings in the input low-resolution simulation results. The second stage uses another CNN to reconstruct small-scale flows between buildings from the output of the first stage, resulting in high-resolution inferences. The CNNs are trained using high-resolution simulation data for the second stage and their coarse-grained version for the first stage as the ground truth, where the high-resolution simulations are conducted independently of the low-resolution simulations used as input. This learning approach separates the spatial scales of inference in each stage. The effectiveness of the proposed method was evaluated using micrometeorological simulations in an actual urban area around Tokyo Station in Japan. The super-resolution simulation successfully inferred high-resolution atmospheric flows, reducing errors by approximately 50\% compared to the low-resolution simulations. Furthermore, the two-stage approach enabled localized high-resolution inferences, reducing GPU memory usage to as low as 12\% during training. The total wall-clock time for 60-min predictions was reduced to 6.83 min, which was 3.32\% of the high-resolution simulation time.
\end{abstract}




\end{frontmatter}


\section{Introduction} \label{sec:introduction}

Global urbanization suggests that various challenges in cities, such as reducing heat stress and energy consumption, will become increasingly important in the future. The estimation of environmental conditions, such as air temperature and wind velocity, is useful for addressing these challenges \citep{Yang+2023BAE}. For instance, architects can incorporate micrometeorological knowledge, such as thermal airflow responses, into building designs to create more comfortable thermal environments \citep{Lam+2021SCS, Hu+2023FAR}. Moreover, if street-scale winds can be predicted in real time, safe drone delivery could be realized \citep{Gianfelice+2022RE}, which would also be cost-effective and environmentally friendly \citep{Chiang+2019AE}. Computational fluid dynamics (CFD) simulations based on physics are essential tools for estimating the atmospheric state \redA{at} microscales in urban areas \citep{Toparlar+2017RSER,Lam+2021SCS}. However, these simulations are computationally expensive due to the need for high spatial resolution, typically a few meters, to resolve buildings and streets \citep{Toparlar+2017RSER, Mirzaei2021SCS}. Reducing this computational burden would make CFD simulations more practical for various urban applications \citep{Mirzaei2021SCS,Hu+2023FAR}.

\redA{Recently, efficient numerical methods based on deep learning have been proposed for simulating urban flow fields \citep[e.g.,][]{Wang+2023UC}. One of the major applications is surrogate modeling, which accelerates simulations by replacing physics-based models with neural networks (NNs). These surrogate models encompass several types of networks,} such as reduced-order models \citep{Xiao+2019BAE, Heaney+2022FP}, graph neural networks \citep{Shao+2023BAE}, and Fourier neural operators \citep{Peng+2024BAE}. These models predict street-scale airflows in a recursive manner, where the predictions for the current time step are used as input for the next step. There are also \redA{surrogate models} that do not conduct predictions over time. For example, using instantaneous sensor data \redA{from streets}, a generative adversarial network can estimate the flow field at that time \citep{Hu+2024BAE}. Additionally, a convolutional neural network (CNN) can estimate the steady airflow from building distributions \citep{Lu+2023GRL}. Apart from these surrogate models, there are also hybrid approaches that combine physics-based and data-driven models, with a typical example being super-resolution.

\redA{Super-resolution (SR), originally a technique for enhancing image resolution, has been applied to accelerate CFD simulations in urban areas \citep[e.g.,][]{Onishi+2019SOLA, Teufel2023}. This acceleration method, referred to as the SR simulation method, has been developed for various fluid systems.}

The SR simulation method consists of two phases: training and testing. In the training phase, low- and high-resolution (LR and HR) physics-based fluid simulations are performed, and the resulting pairs are used as training data. The LR and HR physics-based models may differ only in resolutions \citep{Wang+2021GMD, Wu+2021GRL, Sekiyama+2023AIES}; or they may also differ in the governing equations, such as using different turbulence models \citep{Bao+2022CUAI}. The LR and HR training data have statistical relationships through experimental configurations. For instance, they may be results based on the same initial and boundary conditions \citep{Wang+2020NIPS, Wang+2021GMD}, or the HR simulations may be driven at the lateral boundaries by the LR simulations \citep{Wu+2021GRL, Sekiyama+2023AIES, Teufel2023}. NNs are then trained to infer the HR results from the LR results as input. In the subsequent testing phase (i.e., operational phase), only the LR fluid simulations are conducted, and the results are super-resolved using the trained NNs. These HR inferences can be obtained in a short time because HR numerical integration is not required and the NN inference is typically fast \redA{\citep{Wang+2021GMD, Sekiyama+2023AIES}}.

\begin{figure}[H]
    \centering
    \includegraphics[width=9cm]{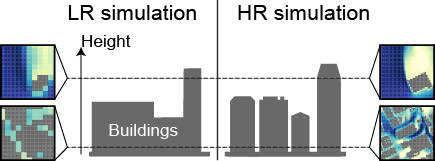}
    \caption{Comparison of building layouts in LR and HR micrometeorological simulations. The colored panels show examples of horizontal snapshots of wind velocity components. \redA{LR and HR are defined as spatial resolutions at which most streets cannot and can be represented, respectively.}}
    \label{fig:schematic-hr-lr-difference}
\end{figure}

The SR in urban areas presents a city-specific challenge: the shapes of buildings, which act as obstacles to airflows, \redA{are resolution-dependent}. At an LR, narrow street canyons between buildings cannot be represented; consequently, the air temperature and wind velocity in these canyons cannot be resolved. In contrast, at an HR, such street-scale flow fields can be resolved (Fig. \ref{fig:schematic-hr-lr-difference}). \redA{In the present study, LR and HR are defined as spatial resolutions at which most streets cannot and can be represented, respectively.} Thus, when enhancing the resolution of flow fields, it is necessary to reconstruct physical quantities in streets. In other words, SR becomes a composite task that includes missing-region reconstruction, which makes the NN training difficult.

\cite{Yasuda+2023BAE} performed this composite-task SR for urban temperatures and winds using an NN that incorporates image-inpainting techniques. Here, image inpainting refers to the reconstruction of missing pixels, and deep learning-based methods have been studied in computer vision \citep{Elharrouss+2020NPL, Qin+2021Display}. In \cite{Yasuda+2023BAE}, the LR input for SR was generated by downsampling HR micrometeorological simulation results. Although this setup is typical in SR studies \citep{Fukami+2023TCFD}, it differs from the SR-simulation setup, where LR data are obtained from separate LR simulations. The accuracy of SR simulations \redA{significantly} decreases if NNs are trained using LR data obtained by downsampling HR simulation results \citep{Wang+2020NIPS, Wang+2021GMD}. This degradation \redA{occurs because the LR input in the test phase is not derived from HR results but} is generated from separate LR simulations. \redA{Given these differences in experimental settings from \cite{Yasuda+2023BAE}, the feasibility of SR simulations in urban areas remains unclear.}

When the LR input is generated from separate LR simulations, the SR process becomes non-local because \redA{LR} flow patterns can differ from those \redA{in} HR simulations. \redA{In this case,} the SR becomes a \redA{multi-resolution} problem: NNs not only infer \redA{HR} flow structures but also modify \redA{LR} flow patterns. \redA{Importantly, these LR patterns have spatially large scales due to their larger grid spacing.} The entire field of physical quantities is useful for recognizing the large-scale patterns. \redA{However,} the data size of the entire field is increased, particularly in three dimensions (3D) due to the additional height dimension. This computational burden has not been obvious in previous \redA{SR-simulation} studies because \redA{they focus on} two-dimensional (2D) flow data, such as precipitation \redA{over cities} \citep{Wu+2021GRL, Teufel2023}. \redA{Therefore, further studies are needed on both the feasibility and computational efficiency of 3D SR simulations in urban areas.}

The objective of the present study is to demonstrate the feasibility of 3D SR simulations for air temperature and wind velocity at street scales. In Section \ref{sec:sr-simulation}, we propose an efficient two-stage SR simulation method that is suitable for \redA{the multi-resolution problem in cities.} To evaluate the proposed method, LR and HR micrometeorological simulations were separately conducted for an actual urban area around Tokyo Station in Japan (Section \ref{sec:methods}). Using these 3D data, we show that the proposed method can achieve about 30 times faster simulations with high accuracy, while reducing the GPU memory consumption (Section \ref{sec:results-discussion}). The conclusions are presented in Section \ref{sec:conclusions}. \redA{Our results} suggest potential for real-time micrometeorology prediction using SR simulation methods.

\section{Two-stage SR simulation method} \label{sec:sr-simulation}

\begin{figure}[H]
    \centering
    \includegraphics[width=16cm]{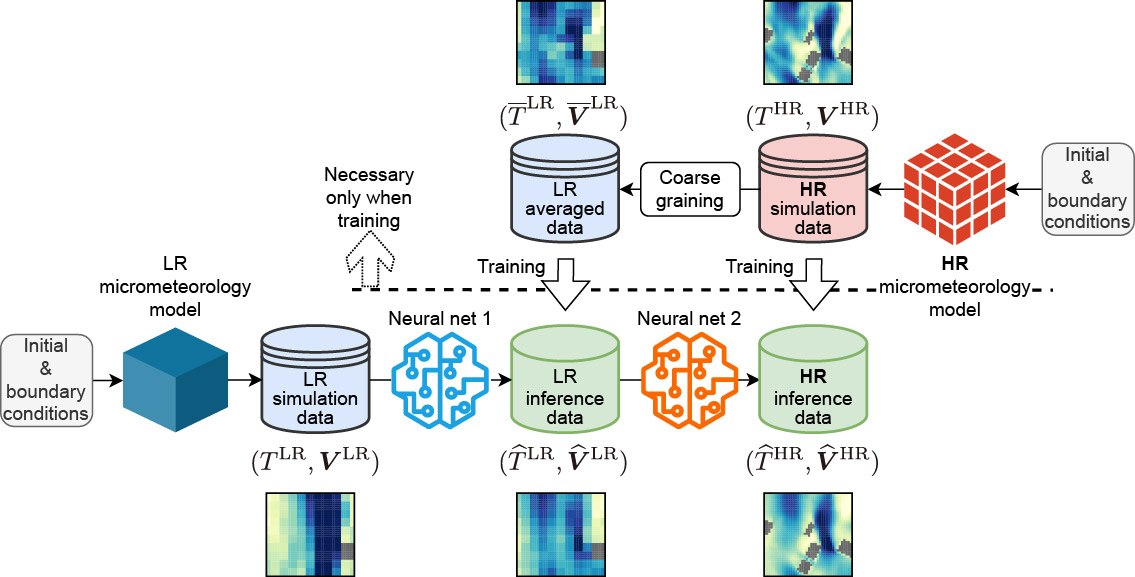}
    \caption{Schematic of the proposed two-stage SR simulation method. Symbols $T$ and $\bm{V}$ represent air temperature and wind velocity, respectively. Detailed definitions of these quantities are given in Section \ref{subsec:data-preparation}. Small colored panels show examples of horizontal snapshots of wind velocity components. The initial and boundary conditions are identical for both LR and HR micrometeorological simulations.}
    \label{fig:sr-simulation}
\end{figure}

The present study proposes a two-stage SR simulation method specifically designed to infer street-scale atmospheric flows in urban areas (Fig. \ref{fig:sr-simulation}). \redA{This two-stage SR method regards the urban SR problem as a multi-resolution problem: the first stage infers the macroscopic flows at LR, while the second stage does the street-scale microscopic flows at HR. This resolution separation would make more accurate inferences than the conventional single-stage SR if the HR and LR grid sizes were limited below the characteristic scales of urban structures, such as city blocks (Section \ref{subsec:dependence-size}).}

For the first stage, an NN is trained for LR inference. Generally, small- and large-scale flows differ between LR and HR micrometeorological simulations, even when both simulations use the same initial and boundary conditions (Fig. \ref{fig:sr-simulation}). HR simulations can represent small-scale turbulence, which likely influences large-scale flows. In urban environments, such turbulence occurs not only through spontaneous time evolution but also due to buildings acting as obstacles. To develop an NN for inferring large-scale flows, we perform coarse-graining of HR simulation results, such as spatial averaging, to generate LR ground-truth data, while LR input data are provided by separate LR simulations. Using these LR input and LR ground truth, the NN is trained to correct large-scale flows without altering the resolution.

In the second stage, the LR inference from the first stage is used as input and super-resolved by another NN (Fig. \ref{fig:sr-simulation}). Ground-truth data for the second stage are directly obtained from the HR simulations. Using these LR input and HR ground truth, the NN is trained to reconstruct small-scale flows in streets. For the composite task of resolution enhancement and missing-region reconstruction, image-inpainting techniques are incorporated into the NN (Section \ref{subsec:neural-nets}).

The ground-truth data, which consist of the HR simulation results and their coarse-grained data, are required only during the training phase. In the operational phase, HR inference can be obtained within the computational time of LR micrometeorological simulations because the NN inference time is usually negligible \redA{\citep{Wang+2021GMD, Sekiyama+2023AIES}}. Thus, the SR simulation system shown in Fig. \ref{fig:sr-simulation} can considerably reduce the total computational time (Section \ref{subsec:eval-hr-inference}).

Another advantage of this two-stage method is its ability to separate spatial scales, thereby reducing computational costs during training. In deep learning, training requires more computational resources than testing because NN computational graphs must be stored in GPU memory for backpropagation. In the first stage, the focus is on large-scale flows; hence, detailed information on small-scale patterns is not essential. The NN can be trained with LR data, which reduces the data volume and facilitates training on the entire spatial region. In the second stage, modification of large-scale flows is no longer required, shifting the focus to the restoration of small-scale flows. Consequently, HR patterns can be inferred from spatially localized inputs; that is, the SR process is localized. This localization allows for training with spatially confined data and reduces the computational burden, even when using HR data (Section \ref{subsec:dependence-size}).

\redA{The proposed method was initially motivated by} image-inpainting studies \citep{Sharma+2018CVPRIPG, Li+2022IEEE}. \redA{Image inpainting is a technique that reconstructs missing pixels.} When filling in large missing regions, long-range correlations \redA{and spatial structures} may be useful \redA{since image inpainting relies on the surrounding pixels.} To efficiently incorporate such long-range information, some image-inpainting NNs use lower-resolution images and reconstruct pixels at the lower resolution \redA{\citep{Sharma+2018CVPRIPG, Li+2022IEEE}}. In our case, the NN in the first stage infers LR atmospheric flows based on entire-region patterns.

\section{Methods} \label{sec:methods}

\subsection{Building-resolving micrometeorological simulations} \label{subsec:micrometeorology}

Building-resolving micrometeorological simulations were conducted using the Multi-Scale Simulator for the Geoenvironment (MSSG) \citep{Onishi+2012JAS, Takahashi+2013JPCS, Sasaki+2016GRL, Matsuda+2018JWEIA}. MSSG is a coupled atmosphere-ocean model that operates on global, meso, and urban scales. The atmospheric component of MSSG, namely MSSG-A, can serve as a building-resolving large eddy simulation (LES) model in conjunction with a 3D radiative transfer model \citep{Matsuda+2018JWEIA}. The governing equations of MSSG-A include the conservation equations of mass, momentum, and energy for compressible flows and the transport equations for the mixing ratios of water substances, such as water vapor. The results of MSSG-A show good agreement with observations under an idealized experimental setup \citep{Matsuda+2018JWEIA}. Further information on the configurations can be found in our previous studies \citep{Matsuda+2018JWEIA, Onishi+2019SOLA}.

\begin{figure}[H]
    \centering
    \includegraphics[width=16cm]{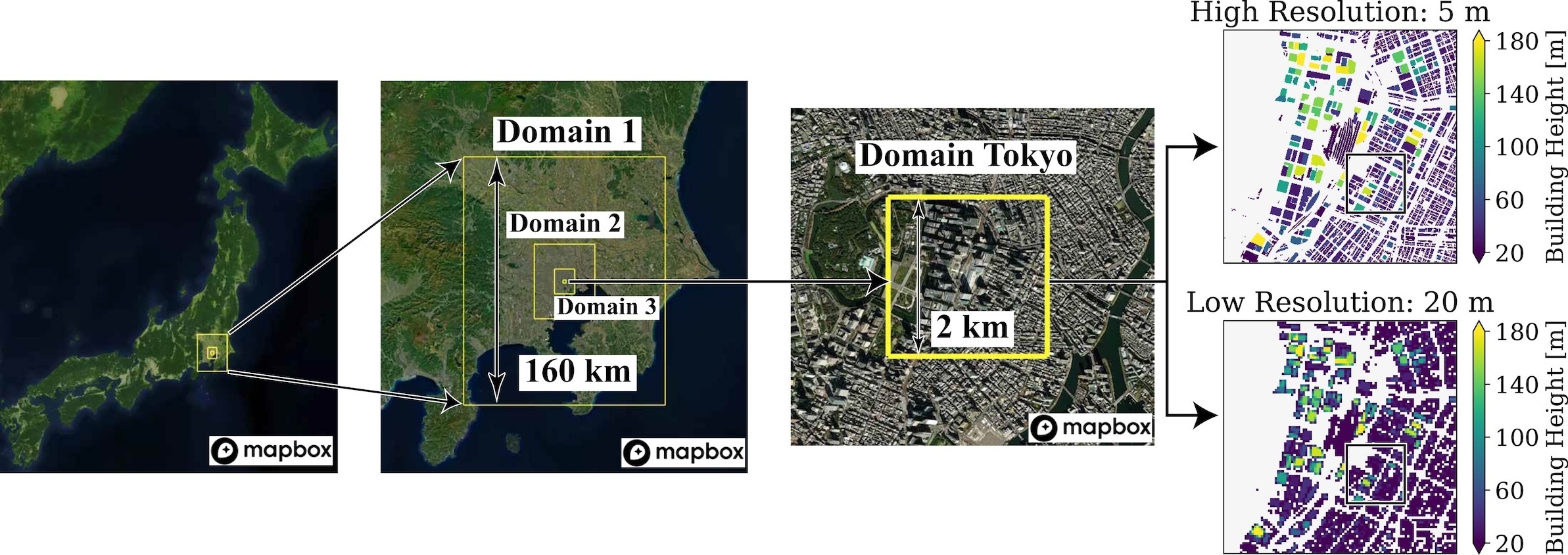}
    \caption{Computational domains for the mesoscale and micrometeorological simulations. The satellite images were obtained from \cite{mapbox} and \cite{openstreetmap}. The micrometeorological simulations were conducted at 5 m and 20 m resolutions; the rightmost panels show the building height distributions used in these simulations. In these distributions, the squares indicate the regions shown in Figs. \redA{\ref{fig:street-width}}, \ref{fig:lr_snapshots}, \ref{fig:hr_snapshots}, and \ref{fig:hr_snapshots_input_dependence}.}
    \label{fig:computation_domains}
\end{figure}

The initial and boundary conditions for SR simulations (Fig. \ref{fig:sr-simulation}) were obtained through mesoscale simulations using MSSG-A. Dynamical downscaling \citep{Giorgi+2015} was performed for an actual urban area around Tokyo Station in Japan (35.6809$^\circ$N and 139.7670$^\circ$E). The mesoscale simulations employed three two-way-coupled nested systems \citep[e.g.,][]{Schlunzen+2011JWEIA}: Domains 1 through 3 (Fig. \ref{fig:computation_domains}). These domains are characterized by horizontal grid points of $160 \times 160$, with grid spacings of $1$ km for Domain 1, $300$ m for Domain 2, and $100$ m for Domain 3. The vertical grids for Domains 1 to 3 are identical and consist of 65 points over an altitude range of $40$ km. The initial and boundary conditions for the mesoscale simulations were obtained from the grid-point-value data of the Meso-Scale Model from the Japan Meteorological Agency \citep{jmbsc}. The resulting data from Domain 3 were used as the boundary and initial conditions for SR simulations.

LR and HR simulation data shown in Fig. \ref{fig:sr-simulation} were obtained via micrometeorological numerical experiments using MSSG-A. The LR and HR simulations employ a $2$ km square domain centered at Tokyo Station (i.e., Domain Tokyo), nested within Domain 3 (Fig. \ref{fig:computation_domains}). Domain Tokyo covers an altitude range of $1.5$ km. The horizontal and vertical grid spacings were set at $20$ m for the LR and $5$ m for the HR. \redA{These values were determined from an analysis of street widths as discussed in the next paragraph.} The initial and boundary conditions for both the HR and LR simulations were provided by the results of Domain 3 \citep[i.e., one-way nesting,][]{Schlunzen+2011JWEIA}. \redA{The bottom boundary has rigid surface conditions with roughness values varying by surface type (concrete, asphalt, or water), while the top boundary is treated as a rigid smooth surface with prognostic variables provided by the results of Domain 3.} The subgrid-scale turbulence model, specifically Smagorinsky-Lilly-type parameterizations, was applied to both the LR and HR simulations \citep{Takahashi+2013JPCS}, \redA{with the Smagorinsky coefficient set to 0.2.} Only the HR simulations incorporated 3D radiative processes \citep{Matsuda+2018JWEIA}, which account for diabatic heating from building surfaces due to solar radiation, as well as the contrast between sunlit and shaded areas caused by buildings. The height distribution of the HR buildings was sourced from an open platform of 3D urban models \citep{PLATEAU2020}. The LR building height distribution was obtained by 2D average pooling from the HR distribution. Specifically, the entire region was divided into $20$ m square blocks, and each block was replaced with the mean value. Although this operation changes the building outlines, it preserves the total volume of the buildings, ensuring that the total atmospheric volume remains unchanged. \redA{The buildings are modeled as rigid-wall objects with a roughness length of 0.005 m for concrete surfaces \citep{Matsuda+2018JWEIA}.}

\begin{figure}[H]
    \centering
    \includegraphics[width=14cm]{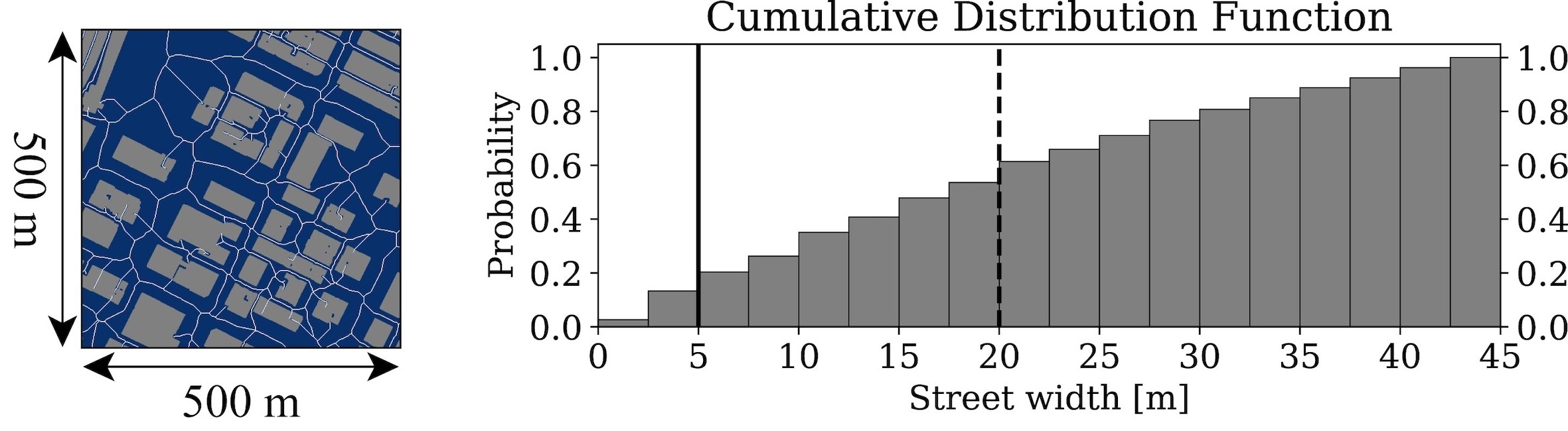}
    \caption{\redA{Street center lines and cumulative distribution function of street widths in Domain Tokyo. The left panel shows a portion of the binary image at a 1-m resolution over the square region marked in Fig \ref{fig:computation_domains}. Blue and gray cells represent streets and buildings, respectively. White lines indicate the street center lines estimated using the method of \cite{Zhang+Suen84ACM}. The right panel shows the cumulative distribution function of street widths. To exclude vacant areas, we restricted the range of widths less than or equal to 45 m; this threshold is the width of the widest street, Sotobori-dori Street. The 5-m width (the HR grid spacing) corresponds to the 14th percentile, indicating that 86\% of the streets are represented at the HR. The 20-m width (the LR grid spacing) corresponds to the 55th percentile, indicating that only 45\% of the streets can be resolved at the LR.}}
    \label{fig:street-width}
\end{figure}

\redA{We examined a histogram of street widths to determine appropriate grid spacings of 20 and 5 m for the LR and HR, respectively. We first binarized a 2D building distribution from the urban model database \citep{PLATEAU2020}. This binary image has a 1-m resolution and takes 0 for building cells and 1 for street cells. Subsequently, the street center lines were estimated using the method of \cite{Zhang+Suen84ACM}. The resultant lines describe the street centers well (Fig. \ref{fig:street-width}). At each center point, we computed the shortest distance to the nearest building cell and doubled it to obtain the full street width. The HR of 5 m was selected because 86\% of the streets are represented at this resolution (Fig. \ref{fig:street-width}). As for LR, previous studies of SR simulations have used a scaling factor from 4 to 10 \citep{Wang+2021GMD, Wu+2021GRL, Sekiyama+2023AIES}. Following them, we set the LR at 20 m (i.e., the scaling factor of 4). At this LR, over half of the streets cannot be resolved (Fig. \ref{fig:street-width}).}

\begin{figure}[H]
    \centering
    \includegraphics[width=9cm]{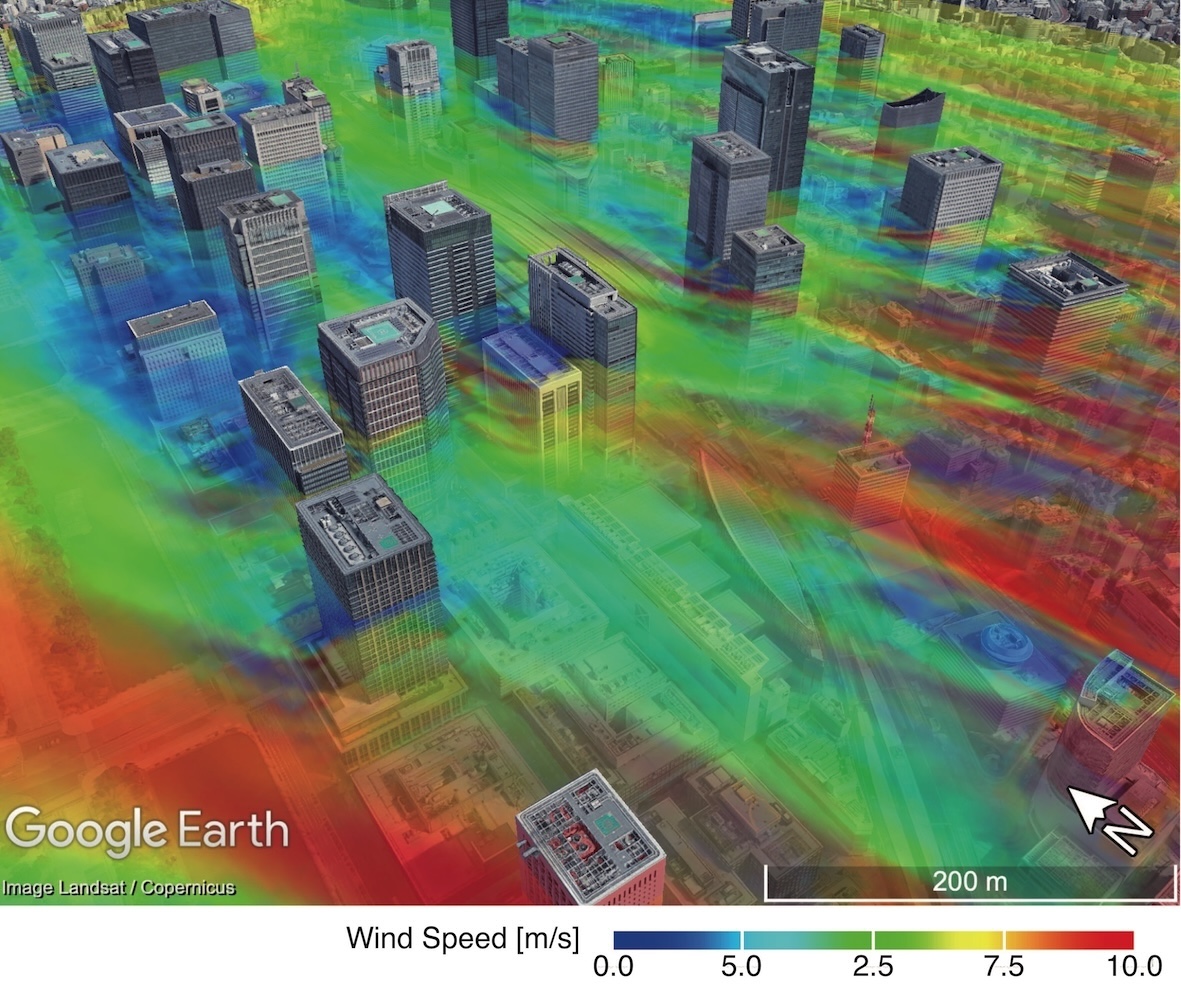}
    \caption{3D distribution of the 1-min averaged wind speed at 2020-08-07 14:45+09:00 obtained from the HR micrometeorological simulation. The wind speed here represents the magnitude of wind velocity. The Volume Data Visualizer for Google Earth (VDVGE) \citep{Kawahara+2015SIGGRAPH} was utilized for the visualization.}
    \label{fig:vdvge}
\end{figure}

Numerical experiments were conducted for hot days between 2013 and 2020. The focus on hot days was due to the potential impact on the risk of heat-related illnesses influenced by the street-level temperature \citep{Kamiya+2019IJERPH}. Another reason for focusing on hot days is to align statistics of all data in the experiments \citep{Guastoni2021}, which is necessary to evaluate model accuracy for complex problems, such as SR simulations in urban areas (Section \ref{sec:introduction}). Specifically, we selected 114 hot summer hours between 2013 and 2020, during which the maximum daily temperature exceeded 35$^\circ$C. \redA{This threshold was defined by the Japan Meteorological Agency and is approximately the 99th percentile of maximum daily temperatures observed during 2010--2017 \citep{Imada+2019SOLA}. In the extracted cases, local wind speeds vary and can reach up to 15 m s$^{-1}$, while their spatial mean is typically a southerly sea breeze of 4--8 m s$^{-1}$ \citep[e.g.,][]{Matsumoto+17JES}. This mean value range (as well as the temperature range) is similar among the 114 cases due to the restriction to hot days.} Figure \ref{fig:vdvge} shows an instantaneous snapshot of the simulated wind speed, where the airflows appear to be blocked by the buildings, and complex turbulent flows develop behind these buildings. Each simulation was carried out for each target hour. The results of the first 30 min were discarded to exclude a statistically non-stationary period due to the initial transient. The remaining 30 min were used to obtain 1-min averaged values. Consequently, 30 snapshots in three dimensions were obtained from each experiment, totaling 3,420 snapshots ($= 30 \times 114$).

\subsection{Data preparation for deep learning} \label{subsec:data-preparation}

We prepared the training data for the two stages (Fig. \ref{fig:sr-simulation}). The HR ground truth for the second stage was obtained from the HR micrometeorological simulations (Section \ref{subsec:micrometeorology}). We extracted the central $1.6$ km square area (i.e., $320 \times 320$ grid points) from the entire $2$ km square domain to eliminate the influence of the lateral damping layer on dynamical downscaling. Vertically, we extracted the bottom $200$ m region (i.e., $40$ grid points) from the entire $1.5$ km height domain, as almost all buildings are lower than $200$ m. In Section \ref{subsec:eval-hr-inference}, we show that the errors decrease with respect to height, suggesting that similar results would be obtained if data from higher than $200$ m were included. Four variables were used in the experiments: air temperature $T^{\rm HR}$ and the eastward, northward, and upward components of wind velocity, denoted as $u^{\rm HR}$, $v^{\rm HR}$, and $w^{\rm HR}$, respectively. Each variable superscripted by ${\rm HR}$ is represented as a 3D numerical array of size $320 \times 320 \times 40$ for the $x$, $y$, and $z$ axes, respectively. The resolution of each array is $5$ m; thus, it covers $1,600$ m in the east-west ($x$-axis) and north-south ($y$-axis) directions and $200$ m in the vertical direction ($z$-axis).

The LR ground truth for the first stage was generated by average pooling from the HR ground truth. Specifically, the HR snapshots were segmented into $20$ m cubes, and each cube was replaced with its average value. In this averaging process, we excluded the grid points within the LR buildings. The resulting data are labeled as $\overline{T}^{\rm LR}$, $\overline{u}^{\rm LR}$, $\overline{v}^{\rm LR}$, and $\overline{w}^{\rm LR}$, where the overlines signify the averaging operation. Each variable is represented as a 3D numerical array of size $80 \times 80 \times 10$ for the $x$, $y$, and $z$ axes, respectively, corresponding to a resolution of $20$ m for each axis.

The input data for the first stage were obtained from the LR micrometeorological simulations (Section \ref{subsec:micrometeorology}). We extracted the central $1.6$ km square and the bottom $200$ m region, similar to the generation of the ground-truth data. The input of $T^{\rm LR}$, $u^{\rm LR}$, $v^{\rm LR}$, $w^{\rm LR}$ is distinct from the set of $\overline{T}^{\rm LR}$, $\overline{u}^{\rm LR}$, $\overline{v}^{\rm LR}$, and $\overline{w}^{\rm LR}$. The former (without overlines) originates directly from the LR simulations, whereas the latter (with overlines) results from applying average pooling to the HR simulation outcomes. Additionally, we input a building mask $B^{\rm LR}$ derived from the LR building height distribution.
\begin{equation}
    B^{\rm LR} = \begin{cases}
        1 & (\text{outside LR buildings}), \\
        0 & (\text{inside LR buildings}).
    \end{cases} \label{eq:building-mask}
\end{equation}
This $B^{\rm LR}$ was created as follows: at each horizontal location, $0$ was assigned to all grid points below the LR building height and $1$ was assigned to the remaining grid points. The LR building height distribution, on which $B^{\rm LR}$ is based, was obtained through 2D average pooling from the HR building height distribution (Section \ref{subsec:micrometeorology}).

The input data for the second stage comprise the LR output of the first stage, along with an HR building mask $B^{\rm HR}$, which was derived from the HR building height distribution in the same manner as in Eq. (\ref{eq:building-mask}). The input, output, and ground truth for both stages are summarized in Table \ref{table:input-output}. The term ``output'' refers to the inferences made by NNs, and these outputs have the same array shapes as those of the corresponding ground truth.

\begin{table}[H]
    \centering
    \caption{Input, output, and ground-truth variables for the first and second stages. Variables with $\rm HR$ ($\rm LR$) are represented as 3D numerical arrays of size $320 \times 320 \times 40$ ($80 \times 80 \times 10$) with $5$ ($20$) m resolution. Hats denote quantities inferred by NNs, whereas variables without hats are quantities from the micrometeorological simulations. The LR variables with overlines were generated through average pooling from the HR simulation results.}
    \label{table:input-output}
    \begin{tabular}{lll}
        \hline
        Stage & Kind & Variables \\
        \hline
        First  & Ground truth & $\overline{T}^{\rm LR}$, $\overline{u}^{\rm LR}$, $\overline{v}^{\rm LR}$, $\overline{w}^{\rm LR}$ \\
        First  & Output       & $\hat{T}^{\rm LR}$, $\hat{u}^{\rm LR}$, $\hat{v}^{\rm LR}$, $\hat{w}^{\rm LR}$ \\
        First  & Input        & $T^{\rm LR}$, $u^{\rm LR}$, $v^{\rm LR}$, $w^{\rm LR}$, $B^{\rm LR}$ \\
        \hline
        Second & Ground truth & $T^{\rm HR}$, $u^{\rm HR}$, $v^{\rm HR}$, $w^{\rm HR}$ \\
        Second & Output       & $\hat{T}^{\rm HR}$, $\hat{u}^{\rm HR}$, $\hat{v}^{\rm HR}$, $\hat{w}^{\rm HR}$ \\
        Second & Input        & $\hat{T}^{\rm LR}$, $\hat{u}^{\rm LR}$, $\hat{v}^{\rm LR}$, $\hat{w}^{\rm LR}$, $B^{\rm HR}$ \\
        \hline
    \end{tabular}
\end{table}

All pairs of the input and ground truth were split into training, validation, and test datasets at ratios of 69\%, 15\%, and 16\%, respectively. This splitting preserves the chronological order to prevent data leakage. Specifically, the training dataset was obtained from 79 micrometeorological simulations between 2013 and 2019 (2,370 pairs); the validation set was obtained from 17 simulations in 2019 (510 pairs); the test set was obtained from 18 simulations in 2020 (540 pairs). \redA{The test dataset is used in Section \ref{sec:results-discussion} to evaluate the model performance with previously unseen data.}

\subsection{Convolutional neural networks (CNNs)} \label{subsec:neural-nets}

\begin{figure}[H]
    \centering
    \includegraphics[width=16cm]{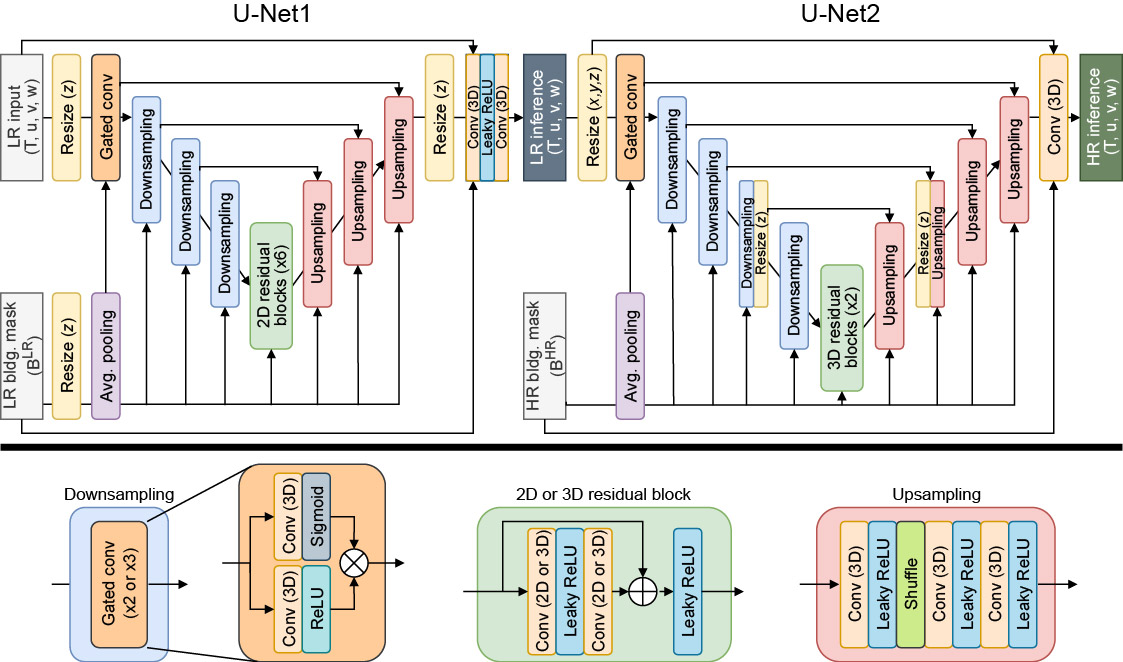}
    \caption{Architecture of U-Net type CNNs for the proposed two-stage SR. ``Conv'' denotes convolution, ``Avg. pooling'' represents 3D average pooling, and ``Shuffle'' refers to the 3D version of pixel shuffle \citep{Shi+2016CVPR}. ``Resize'' indicates nearest-neighbor interpolation, with the following symbols (e.g., $(x,y,z)$) specifying the directions to which the interpolation is applied. The detailed implementation is available in the Zenodo repository (see Data availability).}
    \label{fig:network-architectures}
\end{figure}

We employed two CNNs, U-Net1 and U-Net2, for the first and second stages of the SR, respectively (Fig. \ref{fig:network-architectures}). These CNNs are based on the U-Net architecture proposed in \cite{Yasuda+2023BAE}. U-Net1 infers LR flow fields from LR micrometeorological simulation results; these LR inferences are then super-resolved by U-Net2. We briefly describe U-Net1 and U-Net2, highlighting the differences from the U-Net used in \cite{Yasuda+2023BAE}. \redA{Hyperparameters are described in \ref{sec:hyper-param-appendix}. The full implementations of} U-Net1 and U-Net2 are available in the Zenodo repository (see Data availability). 

U-Net1 and U-Net2 reduce the input size using downsampling blocks and then restore it to its original size through upsampling blocks, similar to other U-Net type NNs \citep{Ronneberger+2015MICCAI,Siddique+2023IEEE}. This process is expected to learn features at various scales. In the downsampling step, we employ an image-inpainting technique called gated convolution \citep{Yu+2019ICCV}, which is effective in reconstructing street-scale flows. Specifically, gated convolution tends to assign greater weights to non-missing values outside buildings using a sigmoid function, thereby reducing the impact of the missing values (i.e., building grids) \redA{and improving accuracy \citep{Yasuda+2023BAE}.} After downsampling, the smaller-sized data are non-linearly transformed through residual blocks \citep{He+2016CVPR}. Subsequently, these data are upsampled to produce output of the same size as the original. Importantly, U-Net2 initially resizes the LR input to the same size as the HR data using nearest-neighbor interpolation, labeled as ``Resize $(x,y,z)$'' in Fig. \ref{fig:network-architectures}, resulting in an HR output. Both U-Net1 and U-Net2 have inputs and outputs consisting of four components: $T$, $u$, $v$, and $w$ (Table \ref{table:input-output}). In addition to these inputs, the LR and HR building masks are fed into U-Net1 and U-Net2, respectively. These building masks are resized by average pooling if necessary and then fed into each block within each network. This technique is effective for the restoration of missing values \citep{Schweri+2021FC}.

The main difference from \cite{Yasuda+2023BAE} is the data size, necessitating resizing. The repeated downsampling and upsampling require the data size to be proportional to $2^n$ for each direction, where $n$ is 3 for U-Net1 and 4 for U-Net2. For the horizontal directions ($x$ and $y$ axes), resizing is not required due to the size of $80$ for the LR and $320$ for the HR (Section \ref{subsec:data-preparation}). However, for the vertical direction ($z$ axis), resizing is needed due to the size of $10$ for the LR and $40$ for the HR. In U-Net2, vertical resizing is inserted after the third downsampling block and before the second upsampling block, labeled as ``Resize $(z)$'' in Fig. \ref{fig:network-architectures}. U-Net1 also uses vertical resizing before downsampling and after upsampling. We confirmed that the accuracy of U-Net1 and U-Net2 is not sensitive to the locations of the resizing blocks.

More importantly, the latent space is 2D for U-Net1, whereas it is 3D for U-Net2. Here, the latent space refers to the feature space where non-linear transformations, labeled as residual blocks, are applied (Fig. \ref{fig:network-architectures}). This two-dimensionality stems from \redA{the small aspect ratios and the definition of LR. The ratio of the vertical size to the horizontal sizes is small,} particularly for the LR data ($80 \times 80 \times 10$). The vertical size of $10$ is reduced to $1$ by downsampling. The two-dimensionalization for the LR data is consistent with the proposed two-stage SR. The first stage of LR inference focuses on large-scale flows above buildings \redA{since most narrow streets cannot be resolved at LR by definition (Section \ref{sec:introduction}), and many grid points are missing at lower levels.} Thus, the vertical relationship between upper and lower flows \redA{is not well represented at LR}. Several studies have reported that temporal information improves the SR of CFD simulations \citep{Xie+2018ACM, Jiang+2020IEEE, Wang+2020NIPS, Bao+2022CUAI, Teufel2023}. When addressing spatio-temporal data, the 2D latent space will become 3D due to the time axis, allowing for the easy application of state-of-the-art techniques from video processing in computer vision \citep{Liu+2022AIR}.

The following preprocessing is applied to treat zero as missing values. Each variable $X$ ($= T$, $u$, $v$, or $w$) is transformed as follows:
\begin{equation}
     {\rm clip}_{[0, 1]}\left( \frac{X-m_X}{s_X}\right). \label{eq:preprocess}
\end{equation}
The clipping function ${\rm clip}_{[0,1]}(x) = \min\{1, \max\{0, x\}\}$ ($x \in \mathbb{R}$) limits the value range to $[0, 1]$. The parameters $m_X$ and $s_X$ are determined such that 99.9\% of $X$ values fall within the range $[0,1]$. The same values of $m_X$ and $s_X$ are used regardless of resolution. After applying Eq. (\ref{eq:preprocess}), the Not a Number (NaN) values at grid points inside buildings are replaced with zero. Due to this preprocessing, the spatial distribution of zero in the input represents grid points inside buildings. This clear interpretation of zero facilitates the learning of 3D image inpainting processes \citep{Yasuda+2023BAE}.

\subsection{Training of the CNNs} \label{subsec:training}

The CNNs, U-Net1 and U-Net2, were trained via supervised learning, following \cite{Yasuda+2023BAE}. The training method is briefly explained here.

The loss function used in training is as follows:
\begin{align}
    \| \bm{{Y} - \bm{\hat{Y}}} \|^2 &+ \lambda_{\rm grd} \| \tilde{B} \odot \nabla(\bm{Y} - \bm{\hat{Y}}) \|^2 \notag \\ &+ \lambda_{\rm div} \| \tilde{B} \odot (\nabla\cdot\bm{V} - \nabla\cdot\bm{\hat{V}}) \|^2, \label{eq:loss}
\end{align}
where $\bm{Y}$ represents the ground truth, $\bm{Y} = (T, u, v, w) = (T, \bm{{V}})$; $\bm{\hat{Y}}$ is the corresponding inference by the CNN; $\lambda_{\rm grd}$ and $\lambda_{\rm div}$ are positive real constants; $\lVert \cdot \rVert^2$ denotes mean squared quantities; $\odot$ represents element-wise multiplication; and $\nabla$ signifies the second-order centered difference. The ground truth and inference are given for the first and second stages, according to Table \ref{table:input-output}. The first term in Eq. (\ref{eq:loss}) evaluates the flow-field values, while the second and third terms quantify the differences in flow-field smoothness and are referred to as the mean gradient error \citep{Schweri+2021FC, Lu+2022SIVP} and the mean divergence error \citep{Schweri+2021FC, Bao+2022CUAI}, respectively. \redA{Note that the velocity field is not divergence-free since MSSG-A solves the compressible flow equations (Section \ref{subsec:micrometeorology}), which is why the difference in divergence is included in the loss function.} The modified building mask $\tilde{B}$ in Eq. (\ref{eq:loss}) differs from the building mask $B$ in Eq. (\ref{eq:building-mask}): $\tilde{B}$ takes values of $0$ at the grid points nearest to buildings as well as inside them, to exclude the invalidity of finite differences near building surfaces.

U-Net1 was initially trained, followed by U-Net2. During U-Net2 training, the inferences from the trained U-Net1 were used as input, \redA{with these inferences stored on disk and loaded when needed.} The Adam optimizer \citep{Kingma+2015ICLR} was employed for both trainings, with learning rates of $5.0 \times 10^{-4}$ for U-Net1 and $3.0 \times 10^{-4}$ for U-Net2. A mini-batch size of $32$ was applied in the training of U-Net1 and U-Net2. The parameters in the loss function [Eq. (\ref{eq:loss})] were set to $\lambda_{\rm grd} = 1$ and $\lambda_{\rm div} = 0$ for U-Net1, and $\lambda_{\rm grd} = 1$ and $\lambda_{\rm div} = 10$ for U-Net2. Each training was terminated by early stopping with a patience parameter of $50$ epochs. \redA{No data augmentation techniques were applied in any training.} During U-Net2 training, $64 \times 64 \times 40$ grid points were randomly cropped from the HR data, and the corresponding region was extracted from the LR input. Since the number of vertical grid points is $40$ at the HR (Section \ref{subsec:data-preparation}), no cropping was performed in the vertical direction. This difference in cropping among directions is due to the strong influence of buildings along the vertical direction \citep{Yasuda+2023BAE}. The sensitivity to the horizontal cropped size for U-Net2 will be discussed in Section \ref{subsec:dependence-size}. For U-Net1 training, the full size of the data was used, and cropping was not applied in any direction due to the smaller data capacity of the LR data.

U-Net1 and U-Net2 were implemented using PyTorch 1.12.1 \citep{pytorch2019} and trained with distributed data parallel using two NVIDIA A100 40GB PCIe GPU boards on the Earth Simulator at the Japan Agency for Marine-Earth Science and Technology (JAMSTEC). The training process took approximately 2 hours for U-Net1 and 10 hours for U-Net2. A single GPU was used for evaluation. Implementation details are available in the Zenodo repository (see Data availability).

\subsection{\redA{Sensitivity test of inference accuracy to spatial size of the training data}} \label{subsec:sensitivity-test-method}

\blueA{In the second stage, snapshots were cropped and only localized data were supplied for training NNs (Section \ref{subsec:training}). If this learning is successful, the inference in the second stage does not strongly depend on the size of} \redA{the cropped training data. We explain here a method to verify this hypothesis.}

\blueA{To examine the accuracy dependence on spatial data size in the second stage, horizontal cropping was applied using one of the following sizes during training: $16 \times 16$, $32 \times 32$, $48 \times 48$, $64 \times 64$, and $80 \times 80$. Vertical cropping was not performed in any case (Section \ref{subsec:training}). For instance, with $32 \times 32$ cropping, each HR training data sample has a size of $32 \times 32 \times 40$ for the $x$, $y$, and $z$ axes, respectively. For comparison, we trained an NN that has the same architecture as U-Net2 but uses the LR simulation results as input. We refer to the SR by this NN as single-stage SR. Compared to the second-stage SR, the single-stage SR needs to modify large-scale flows and restore small-scale flows simultaneously.} \redA{The GPU memory consumption of the single-stage training is equivalent to that of the second-stage training, as both use U-Net2 to super-resolve one LR snapshot of the same input size.} \blueA{To investigate inference sensitivity, we varied the pseudo-random seed used for the weight initialization \citep{He+2015ICCV} and conducted five sets of training for each setup. During testing, the full-size data without cropping were input to assess inference accuracy.}

\subsection{Metrics for evaluation of the CNNs} \label{subsec:evaluation}

The CNN performance was evaluated using two types of metrics: pointwise accuracy and pattern consistency. \redA{These two types are frequently used in SR studies \citep{Chauhan+2023IEEE, Lepcha+2023IF} because they complement each other. The pointwise error quantifies local differences but may overestimate discrepancies when flow patterns are correctly predicted with a slight shift. The pattern consistency mitigates this effect by evaluating structural similarity between inferred and ground-truth patterns, although it may overlook significant local deviations.}

The pointwise errors for air temperature and wind velocity are defined as follows:
\begin{align}
    \left\lvert\Delta T\right\rvert &= \frac{1}{N} \sum_{\text{const height}}  B\left\lvert T - \hat{T} \right\rvert, \notag \\
    &= \frac{1}{N} \sum_{\text{const height}}  B\sqrt{\left(T - \hat{T}\right)^2}, \label{eq:mae-T}
\end{align}
and
\begin{align}    
    \left\lvert\Delta \bm{V}\right\rvert &= \frac{1}{N} \sum_{\text{const height}}  B \left\lvert \bm{V} - \bm{\hat{V}} \right\rvert, \notag \\
    &= \frac{1}{N} \sum_{\text{const height}}  B \sqrt{\left(u - \hat{u}\right)^2 + \left(v - \hat{v}\right)^2 + \left(w - \hat{w}\right)^2 }, \label{eq:mae-V}
\end{align}
where $\bm{V}$ denotes the wind velocity, quantities without hats represent the ground-truth variables, the summation is taken at a constant height over all test data, and $N$ denotes the number of grid points outside the buildings (i.e., $N = \sum_{\text{const height}} B$). Equations (\ref{eq:mae-T}) and (\ref{eq:mae-V}) are referred to as the error norms for temperature and velocity, respectively. These norms were applied to the LR and HR variables in the first and second stages, respectively, according to Table \ref{table:input-output}.

Pattern consistency is estimated using the mean structural similarity index measure (MSSIM) \citep{Wang+2004IEEE}. This metric is determined based on first- and second-moment quantities, such as covariance:
\begin{align}
    \text{MSSIM loss} &= 1 - {\rm MSSIM}, \notag \\ &= 1 - \frac{1}{N}\sum_{\text{const height}} \left[\frac{\left(2\mu\hat{\mu} + {\rm C_1} \right) \left(2\chi + {\rm C_2} \right)}{\left(\mu^2 + \hat{\mu}^2 + {\rm C_1}\right) \left( \sigma^2 + \hat{\sigma}^2 + {\rm C_2} \right)} \right], \label{eq:mssim}
\end{align}
where ${\rm C_1} = 0.01$ and ${\rm C_2} = 0.03$. Here, $\mu$ and $\sigma^2$ denote the mean and variance of the ground truth, respectively, $\hat{\mu}$ and $\hat{\sigma}^2$ are the corresponding quantities of the inference, and $\chi$ represents the covariance between the ground truth and inference. The MSSIM loss, with values greater than or equal to 0, quantifies the similarity between the spatial patterns of the inference and the ground truth; smaller values indicate higher similarity. \redA{MSSIM values above 0.9 indicate perceptually similar patterns \citep{Wang+2004IEEE}, suggesting that an MSSIM loss around 0.1 represents acceptable performance.} The MSSIM loss was computed separately for $T$, $u$, $v$, and $w$, then averaged across these four components. We confirmed that the differences in MSSIM loss among the components are sufficiently small ($\lesssim 0.05$). A detailed discussion of MSSIM can be found in previous studies \citep{Wang+2004IEEE}.

\section{Results and discussion} \label{sec:results-discussion}

\subsection{Evaluation of LR inferences} \label{subsec:eval-lr-inference}

We first demonstrate the accuracy of the LR inference by U-Net1. In the first stage, the ground-truth LR data are obtained by averaging the results of the HR micrometeorological simulations, while the input comes directly from the LR simulation results. U-Net1 modifies the flow patterns in the LR simulations to more closely resemble those of the ground truth.

\begin{figure}[H]
    \centering
    \includegraphics[width=14.55cm]{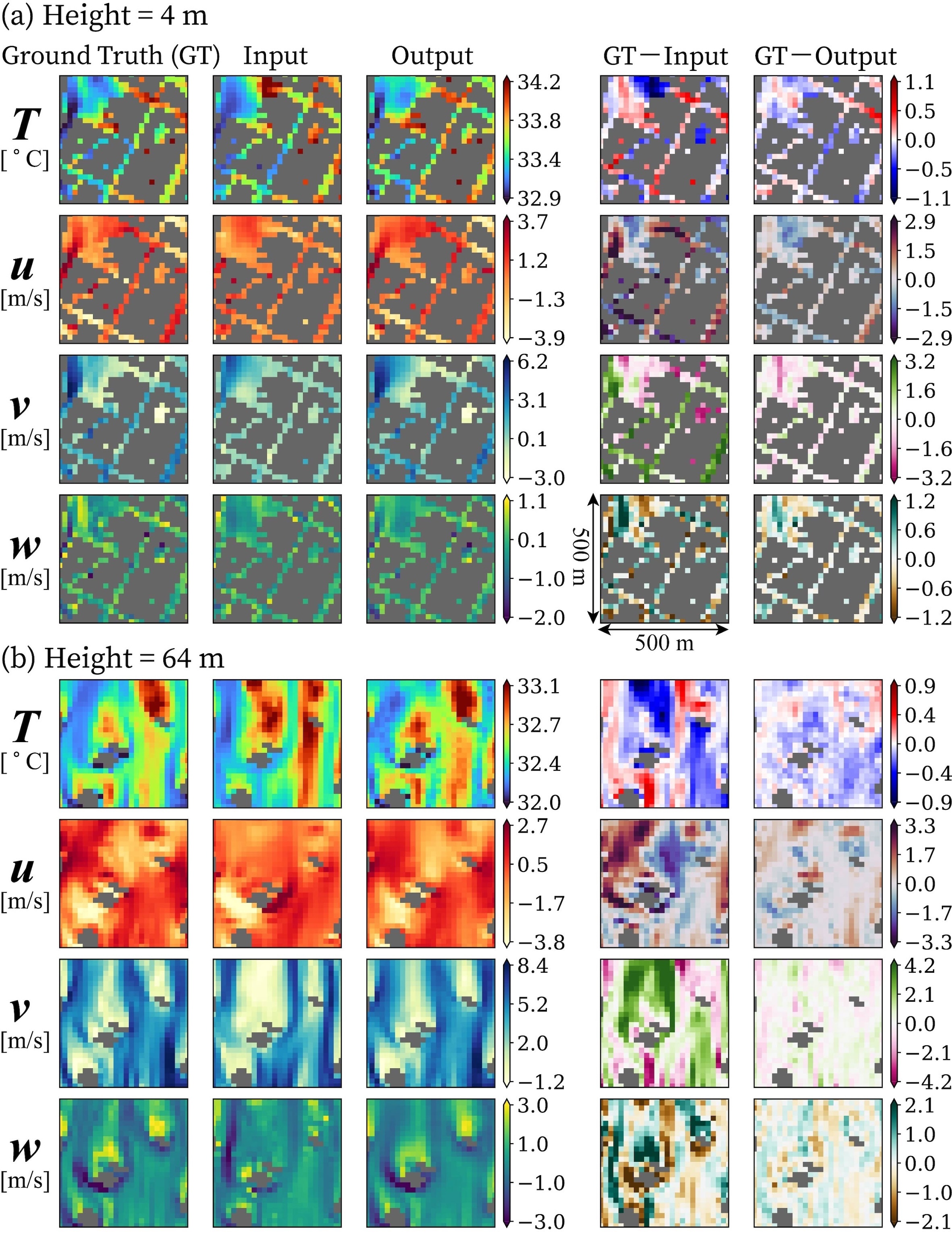}
    \caption{Example of LR inference at heights of (a) $4$ m and (b) $64$ m: air temperature ($T$) and eastward, northward, and upward components of wind velocity ($u$, $v$, and $w$, respectively). Building cells are colored gray. \redA{The two columns on the right display the differences between the ground truth (GT) and both input and output.} The region shown here corresponds to the square marked in the building height distribution at $20$ m resolution in Fig. \ref{fig:computation_domains}.}
    \label{fig:lr_snapshots}
\end{figure}

Figure \ref{fig:lr_snapshots} shows an example of inference snapshots at heights of $4$ and $64$ m above the ground, with gray areas indicating building cells. At $4$ m, U-Net1 infers a flow pattern that blows over the street toward the vacant area in the upper left (Fig. \ref{fig:lr_snapshots}a). \redA{Although U-Net1 reduces the error magnitudes, the differences between the input and inference are not very distinct} across the entire snapshot, as many narrow streets are not represented due to the coarse resolution. In contrast, at $64$ m, where there are fewer buildings, the differences between the input and inference become more distinct (Fig. \ref{fig:lr_snapshots}b). U-Net1 modifies the flow patterns behind the buildings, making them more similar to the ground truth. \redA{These modifications lead to substantial error reduction.}

\begin{figure}[H]
    \centering
    \includegraphics[width=16cm]{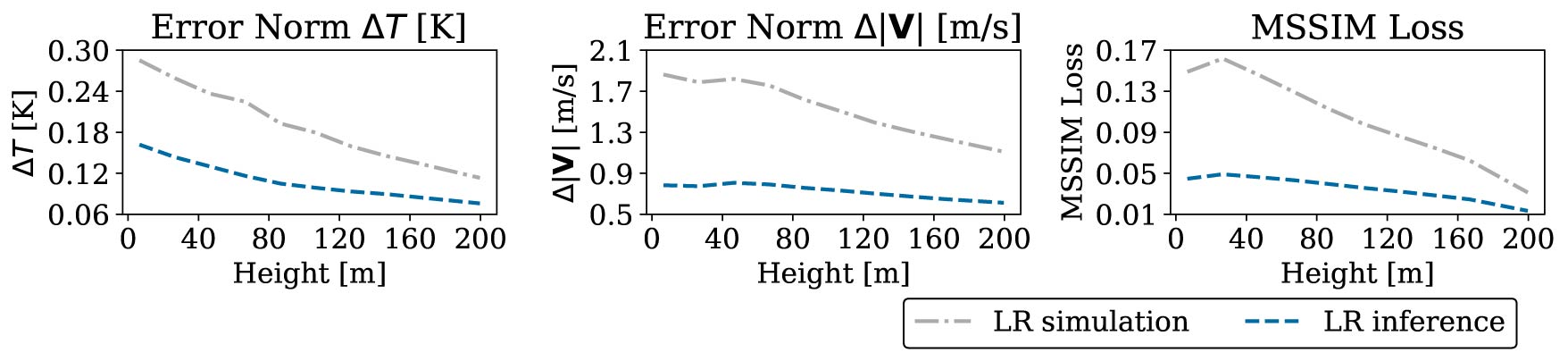}
    \caption{Height dependence of the test metrics for the LR inference. The error norms and MSSIM losses are defined in Eqs. (\ref{eq:mae-T}), (\ref{eq:mae-V}), and (\ref{eq:mssim}). The LR micrometeorological simulation results (dash-dotted lines) are the input to U-Net1, while its output is labeled as LR inference (dashed lines).}
    \label{fig:lr_height_dependence}
\end{figure}

Quantitative analysis further confirms the accuracy of U-Net1 inference. Figure \ref{fig:lr_height_dependence} shows the height dependence of the test errors, averaged over all time steps. The error reduction with increasing height may reflect the diminishing influence of buildings. At all heights, the errors in the LR inference are lower than those in the LR micrometeorological simulation. At the $64$ m height, where many LR buildings are below this level (Fig. \ref{fig:lr_snapshots}), the temperature error decreases from $0.22$ to $0.12$ K, and the velocity error reduces from $1.8$ to $0.79$ m s$^{-1}$ (both approximately halved). The MSSIM loss drops from $0.13$ to $0.043$ (about one-third).

\begin{figure}[H]
    \centering
    \includegraphics[width=16cm]{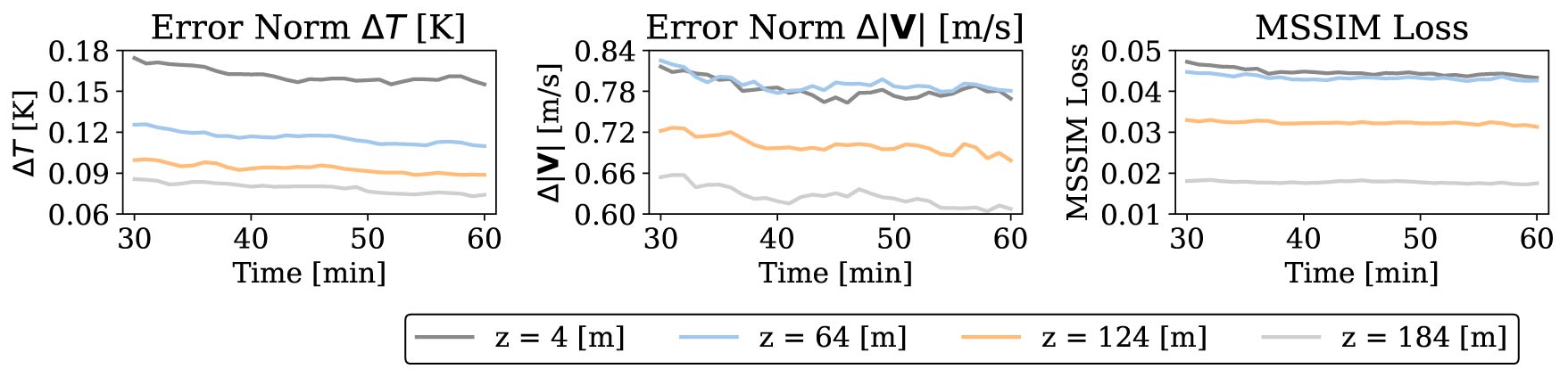}
    \caption{Time dependence of the test metrics for the LR inference. The error norms and MSSIM losses are defined in Eqs. (\ref{eq:mae-T}), (\ref{eq:mae-V}), and (\ref{eq:mssim}).}
    \label{fig:lr_time_dependence}
\end{figure}

Figure \ref{fig:lr_time_dependence} shows time series of the test errors for the LR inference at various heights, where averaging was performed at each time step across all test simulations. The data from 30 to 60 min were utilized (Section \ref{subsec:micrometeorology}); hence, the time axes in the figure start at 30 min. The test errors tend to decrease with height, as shown in Fig. \ref{fig:lr_height_dependence}. The velocity error norm and MSSIM loss exhibit similar magnitudes between heights of $4$ and $64$ m, likely because it is more difficult to reduce error metrics based on multiple components (Section \ref{subsec:eval-hr-inference}). Notably, the errors remain nearly constant over time, with slight reductions in temperature and velocity errors. This result suggests an advantage of SR simulations: time evolution is computed by the LR physics-based model, not by data-driven models, potentially suppressing error growth over time. Previous SR-simulation studies suggest that errors tend not to grow over time \citep{Sekiyama+2023AIES, Teufel2023}, consistent with our result.

\subsection{Evaluation of HR Inferences} \label{subsec:eval-hr-inference}

We next demonstrate the accuracy of HR inference by U-Net2. In this second stage, the ground-truth HR data are directly obtained from the HR micrometeorological simulations, while the input data are the LR inferences from U-Net1 in the first stage. U-Net2 reconstructs small-scale flows in street canyons.

\begin{figure}[H]
    \centering
    \includegraphics[width=14.55cm]{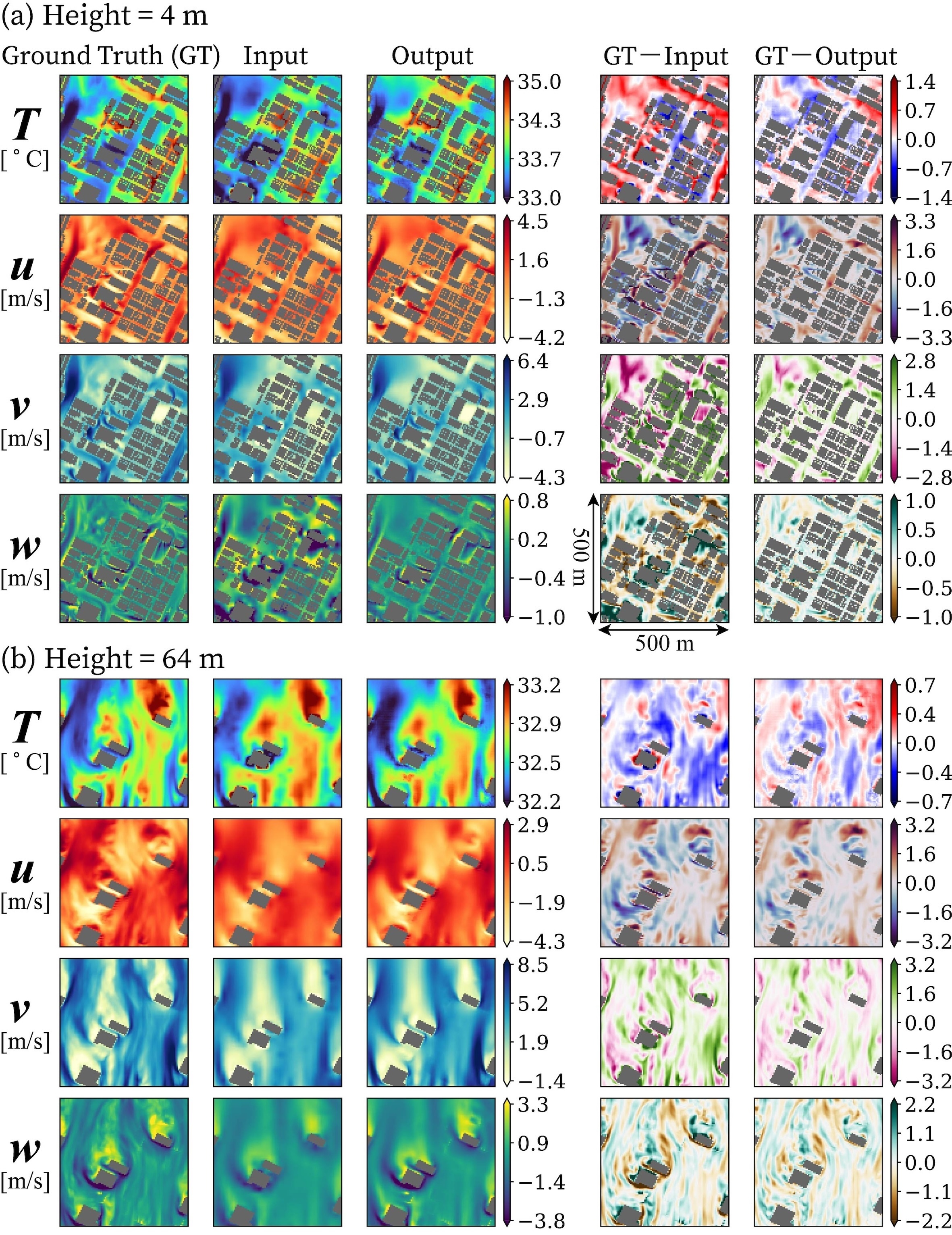}
    \caption{Example of HR inference at heights of (a) $4$ m and (b) $64$ m: air temperature ($T$) and eastward, northward, and upward components of wind velocity ($u$, $v$, and $w$, respectively). Building cells are colored gray. \redA{The two columns on the right display the differences between the ground truth (GT) and both input and output.} The region shown here corresponds to the square marked in the building height distribution at $5$ m resolution in Fig. \ref{fig:computation_domains}.}
    \label{fig:hr_snapshots}
\end{figure}

Figure \ref{fig:hr_snapshots} shows an example of inferences at heights of $4$ and $64$ m, with gray areas indicating building cells. The narrow streets between the buildings are represented at a resolution of $5$ m, in contrast to the lower resolution of $20$ m. For comparison, the street-scale flows in the input shown in Fig. \ref{fig:hr_snapshots}a were estimated by applying linear extrapolation to the LR inference shown in Fig. \ref{fig:lr_snapshots}. At $4$ m, U-Net2 successfully reconstructs the temperature and velocity (Fig. \ref{fig:hr_snapshots}a). In particular, \redA{the large errors between buildings are substantially reduced, and the local patterns in air temperatures and winds become more consistent with those in the ground truth.} At $64$ m, the input and output exhibit similar patterns, as expected (Fig. \ref{fig:hr_snapshots}b), \redA{although the pattern magnitudes in the output are slightly adjusted to better match those of the ground truth.}

\begin{figure}[H]
    \centering
    \includegraphics[width=16cm]{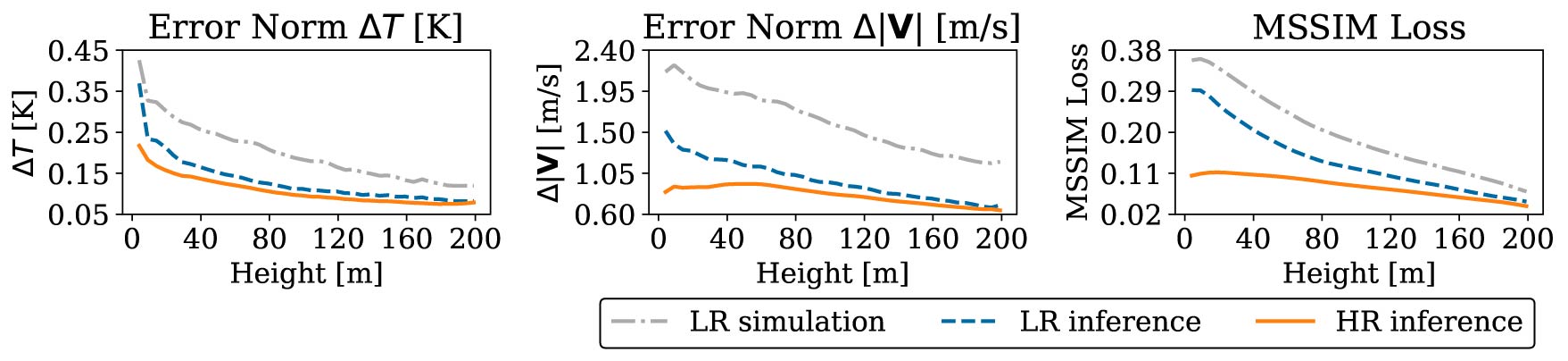}
    \caption{Height dependence of the test metrics for the LR micrometeorological simulation, LR inference, and HR inference. The error norms and MSSIM losses are defined in Eqs. (\ref{eq:mae-T}), (\ref{eq:mae-V}), and (\ref{eq:mssim}). The LR micrometeorological simulation result is the input to U-Net1. The output of U-Net1 (LR inference) serves as input to U-Net2. The final output from U-Net2 is labeled HR inference.}
    \label{fig:hr_height_dependence}
\end{figure}

Quantitative analysis further confirms the accuracy of U-Net2 inference. Figure \ref{fig:hr_height_dependence} compares the height dependence of the test errors among the LR micrometeorological simulation (dash-dotted lines), LR inference (dashed lines), and HR inference (solid lines). As in Fig. \ref{fig:hr_snapshots}, linear extrapolation was used to estimate street-scale flows in the LR simulation and LR inference. The ground truth used here differs from that in Fig. \ref{fig:lr_height_dependence}. In Fig. \ref{fig:hr_height_dependence}, we used the original HR data from the micrometeorological simulations as the ground truth, whereas in Fig. \ref{fig:lr_height_dependence}, their coarse-grained version was used. This distinction leads to larger errors for the LR simulation and LR inference near the ground (dash-dotted and dashed lines, respectively), compared to those in Fig. \ref{fig:lr_height_dependence}. Across all heights, the HR inference (solid lines) shows the smallest test errors (Fig. \ref{fig:hr_height_dependence}). The difference in test errors is most evident near the ground and tends to decrease with height, likely reflecting the greater impact of buildings at lower heights.

\redA{U-Net2 exhibits a tendency to slightly underestimate the absolute values of all physical quantities, regardless of height and distance from building surfaces. This underestimation likely stems from the use of average voxel-wise loss in Eq. \ref{eq:loss} during training. When minimizing such averaged losses, NNs are encouraged to infer values closer to the mean to avoid large errors. This behavior is a well-known characteristic of NNs for SR, and various approaches, such as adversarial networks, have been proposed to mitigate this behavior \citep{Lepcha+2023IF}.}

\begin{figure}[H]
    \centering
    \includegraphics[width=14cm]{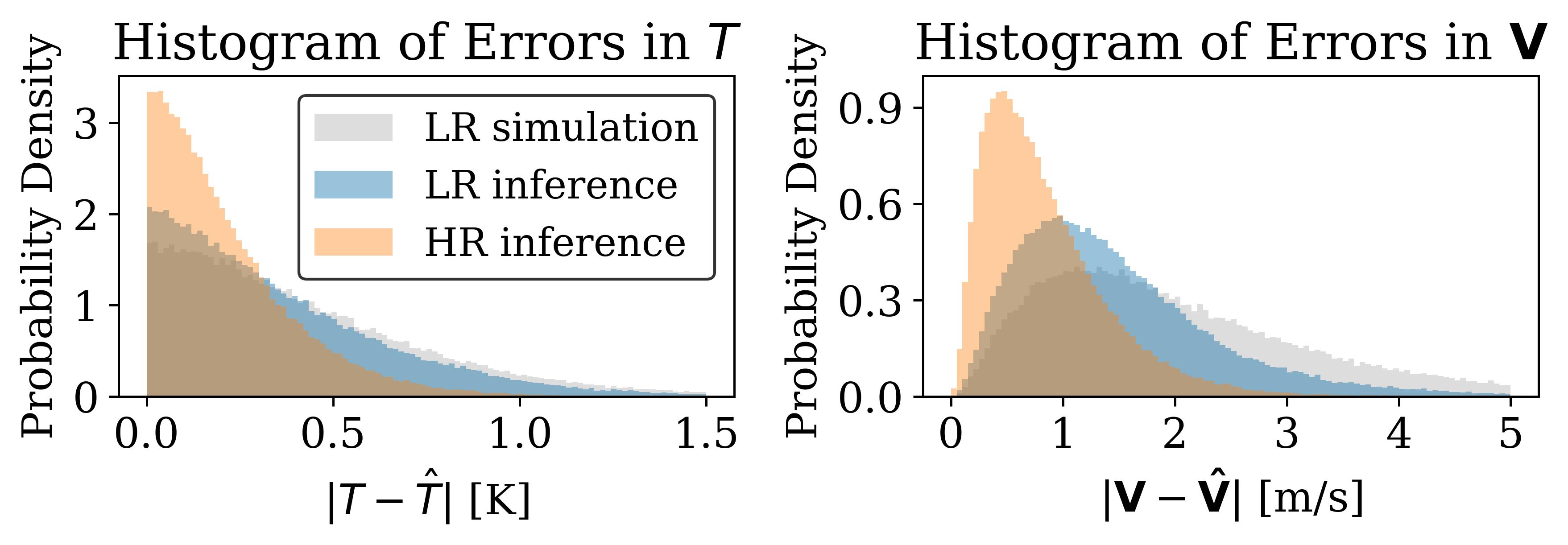}
    \caption{Histograms of pointwise errors in temperature and velocity at $4$ m height. The temperature error is $\left\lvert T - \hat{T} \right\rvert$ in Eq. (\ref{eq:mae-T}). The velocity error is $\left\lvert \bm{V} - \bm{\hat{V}} \right\rvert$ in Eq. (\ref{eq:mae-V}). Each histogram contains 100,000 randomly sampled error values from all test results.}
    \label{fig:hr_histograms}
\end{figure}

To further examine the test errors near the ground, Fig. \ref{fig:hr_histograms} compares the histograms of pointwise errors from the LR simulation, LR inference, and HR inference at $4$ m (the lowest level). These pointwise errors are the quantities in the summations of Eqs. (\ref{eq:mae-T}) and (\ref{eq:mae-V}). The histogram for temperature error peaks near zero, whereas the histogram for velocity error peaks away from zero (Fig. \ref{fig:hr_histograms}). This difference in histogram shapes, also observed in \cite{Yasuda+2023BAE}, likely stems from the error definitions: the temperature error is computed as the absolute difference [Eq. (\ref{eq:mae-T})], whereas the velocity error is defined as the sum of squared differences [Eq. (\ref{eq:mae-V})]. This definition suggests that all velocity components need to be reduced to decrease the velocity error norm, likely making error reduction more difficult. Notably, the histograms for HR inferences are the narrowest. Indeed, both the average values and the $95$th percentiles for HR inferences are approximately 50\% of those for LR simulations and 60\% of those for LR inferences, as reported in Table \ref{table:95-percentiles}.

\begin{table}[H]
    \centering
    \caption{Average and 95th percentile values for pointwise errors in temperature and velocity, derived from the histograms in Fig. \ref{fig:hr_histograms}.}
    \label{table:95-percentiles}
    \begin{tabular}{llll}
        \hline
        Kind & Error metric & Average value & $95$th percentile value \\
        \hline
        LR simulation& $\rvert T^{\rm HR} - {T}^{\rm LR} \lvert$      & $0.426$ K & $1.11$  K \\
        LR inference & $\rvert T^{\rm HR} - \hat{T}^{\rm LR} \lvert$           & $0.369$ K & $0.986$ K \\
        HR inference & $\rvert T^{\rm HR} - \hat{T}^{\rm HR} \lvert$           & $0.217$ K & $0.590$ K \\
        \hline
        LR simulation& $\rvert \bm{V^{\rm HR}} - \bm{{V}^{\rm LR}} \lvert$ & $2.16$ m s$^{-1}$ & $4.99$ m s$^{-1}$ \\
        LR inference & $\rvert \bm{V^{\rm HR}} - \bm{\hat{V}^{\rm LR}} \lvert$ & $1.51$ m s$^{-1}$  & $3.32$ m s$^{-1}$ \\
        HR inference & $\rvert \bm{V^{\rm HR}} - \bm{\hat{V}^{\rm HR}} \lvert$ & $0.845$ m s$^{-1}$ & $1.94$ m s$^{-1}$ \\
        \hline
    \end{tabular}
\end{table}

The time series of the test errors for HR inferences were similar to those in Fig. \ref{fig:lr_time_dependence}, indicating nearly constant error over time (not shown in detail). This result confirms an advantage of SR simulations again: the error accumulation can be suppressed due to the time evolution by physics-based models.

Finally, we discuss the total inference time for 60-min predictions. We report wall-clock times based on measurements taken on the Earth Simulator at JAMSTEC, which employs AMD EPYC 7742 (CPUs) and NVIDIA A100 (GPUs). HR simulations used 256 CPU cores, LR simulations used 40 CPU cores, and SR inferences used a single GPU board. On average, HR micrometeorological simulations required 206 min. LR micrometeorological simulations took 6.19 min. The SR process for the 60-min LR data (60 snapshots) totaled 38.9 s, with 1.43 s for initial LR inferences and 37.5 s for subsequent HR inferences. Consequently, the SR simulation system completed 60-min predictions in 6.83 min on average, reducing the HR computation time from 206 min to 3.32\% (a 30 times speedup). This speedup factor is comparable with those reported in the latest surrogate modeling approaches for urban 3D flow simulations \citep{Shao+2023BAE, Peng+2024BAE}, although direct comparisons are difficult due to the different experimental setups.

There is potential for further reduction in computation time. Most of the total inference time comes from integrating the LR micrometeorology model. Generally, lower-resolution models reduce computation time more. The present study uses a four-fold LR, whereas \cite{Yasuda2022BAE} suggests that an eight-fold LR can be employed. Investigating the use of lower-resolution models is an important future task.

\subsection{Dependence on spatial data size in training} \label{subsec:dependence-size}

\redA{We demonstrate that the inference in the second stage requires only local patterns and allows us to perform training on cropped data. This approach is particularly effective when the spatial size of training data is limited below the characteristic scales of city blocks. Note that the evaluation method used here is explained in Section \ref{subsec:sensitivity-test-method}.}

\begin{figure}[H]
    \centering
    \includegraphics[width=16cm]{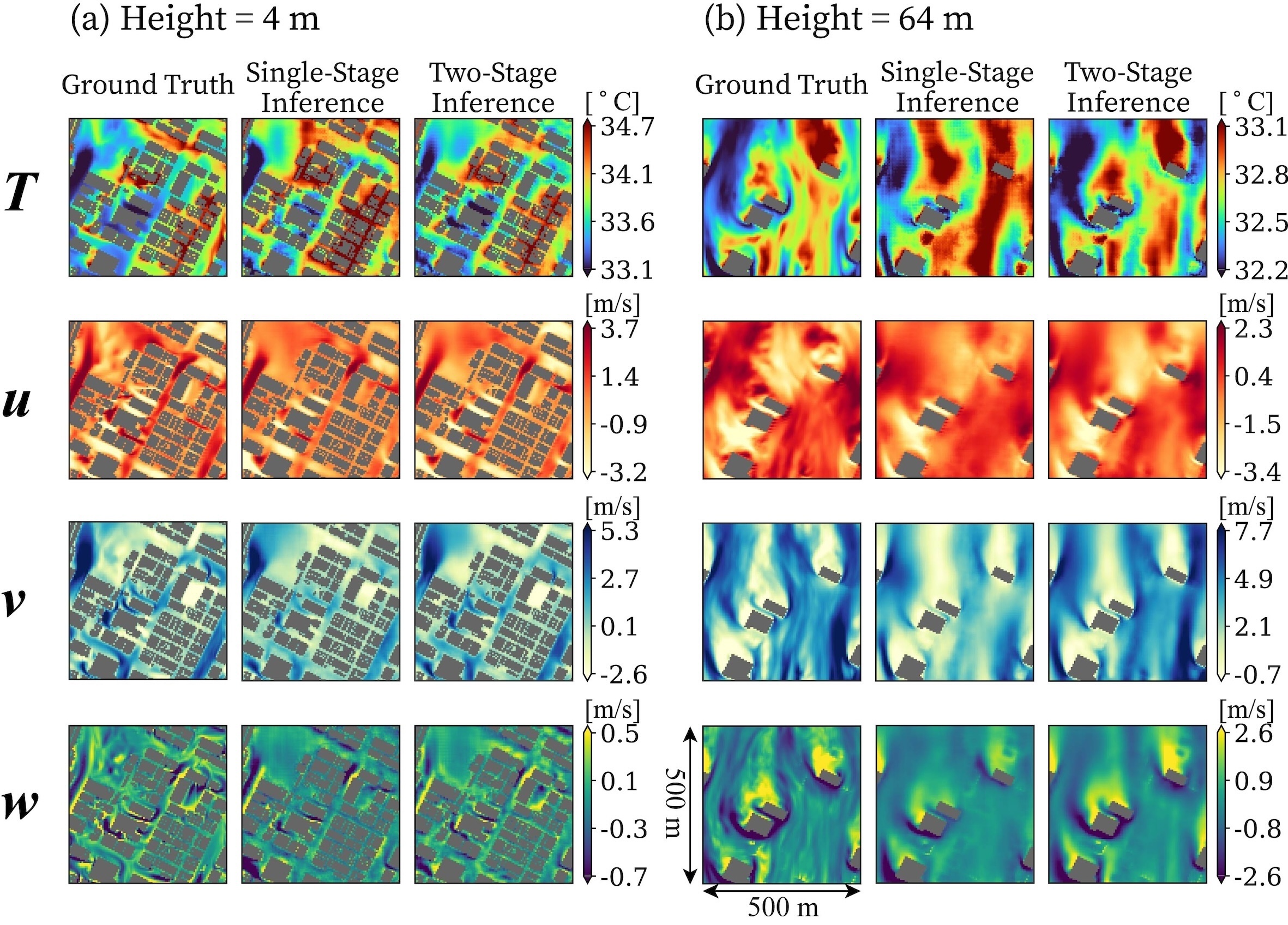}
    \caption{Comparison of HR inferences between the single- and two-stage SR at heights of (a) $4$ m and (b) $64$ m: air temperature ($T$) and eastward, northward, and upward components of wind velocity ($u$, $v$, and $w$, respectively). During training, horizontal cropping of $16\times16$ was applied. \redA{This HR size of $16 \times 16$ corresponds to the LR input size of $4 \times 4$.} Building cells are colored gray. The region shown here corresponds to the square marked in the building height distribution at $5$ m resolution in Fig. \ref{fig:computation_domains}.}
    \label{fig:hr_snapshots_input_dependence}
\end{figure}

Figure \ref{fig:hr_snapshots_input_dependence} compares snapshots of the single- and two-stage SR inference \redA{when the spatial size of the training data is sufficiently small, namely $16 \times 16$.} At a height of $4$ m, both SR methods reconstruct strong local winds between buildings, although the single-stage inference slightly underestimates wind intensity. This result indicates that fine-scale flow reconstruction can be learned from spatially localized data. At $64$ m, where flow scales are larger due to fewer buildings, the single-stage inference shows different flow patterns from those of the ground truth. In particular, the high-temperature and weak-wind areas are wider in the $T$ and $v$ snapshots, respectively. This result suggests that data of insufficient spatial size lead to failure in learning large-scale flow corrections. In contrast, the two-stage inference maintains high accuracy and yields similar patterns at $64$ m, even when using small-size data.

\begin{figure}[H]
    \centering
    \includegraphics[width=16cm]{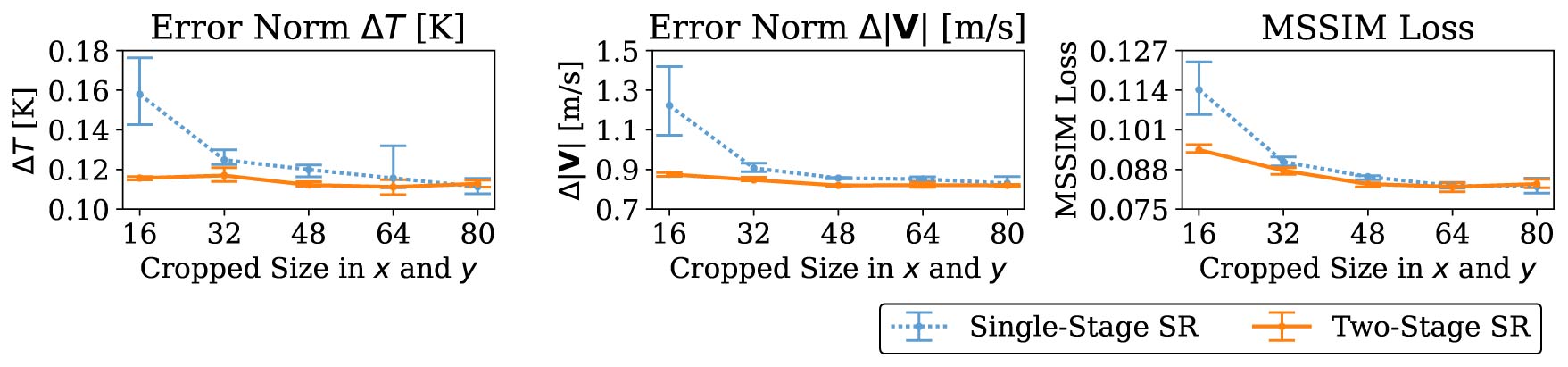}
    \caption{Dependence of the test errors on cropped data sizes. Horizontal cropping was applied only in training U-Net2. The error norms and MSSIM losses are defined in Eqs. (\ref{eq:mae-T}), (\ref{eq:mae-V}), and (\ref{eq:mssim}). For each size, the errors were averaged for U-Net2 using five different pseudo-random seeds in the weight initialization. The averaged scores are indicated by lines, while the error bars show the maximum and minimum values for the five seeds.}
    \label{fig:hr_size_dependence}
\end{figure}

Figure \ref{fig:hr_size_dependence} shows the dependence of the test errors on cropped sizes. For each size, the test errors were averaged for the five U-Net2s trained using different pseudo-random seeds. The averages over the five scores are indicated by lines, while the error bars show the maximum and minimum values. Compared to the single-stage SR, the two-stage SR exhibits less dependence on cropped size. In particular, the temperature and velocity error norms are nearly constant across cropped sizes. Although the MSSIM loss tends to decrease with increasing size, this tendency is much weaker in the two-stage SR. For the single-stage SR, all test errors decrease strongly as the size changes from $16$ to $32$. At the size of $16$, U-Net2 fails to learn how to correct large-scale flows (Fig. \ref{fig:hr_snapshots_input_dependence}). When the size is equal to or greater than $32$, the difference in accuracy between the single- and two-stage SR becomes small. This result implies that U-Net2 can simultaneously learn to correct large-scale flows and reconstruct small-scale flows from input data of sufficient size. Importantly, the error bars are quite large not only at $16$ but also at $64$ for the single-stage SR (see the temperature error norm in Fig. \ref{fig:hr_size_dependence}), suggesting that such simultaneous learning is more challenging than separate, two-stage learning. These results indicate that the two-stage SR allows for training with smaller spatial size data, which not only localizes but also stabilizes HR inference processes.

\begin{figure}[H]
    \centering
    \includegraphics[width=9cm]{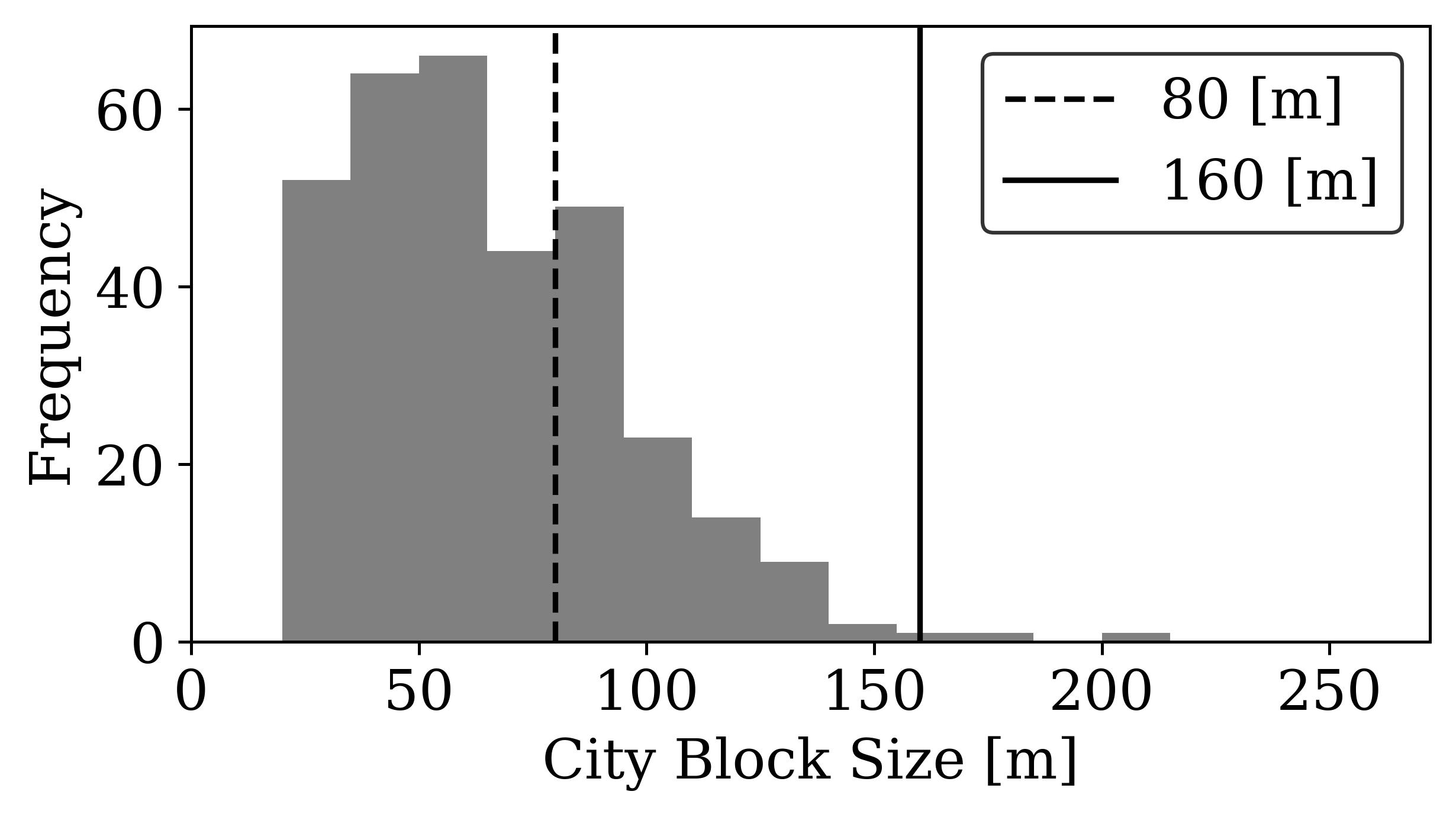}
    \caption{Histogram of city block sizes. Each city block is approximated as a rectangle, and the average length of its width and height is used to create the histogram. The 80 m and 160 m sizes correspond to 16 and 32 grid points \redA{at the HR}, respectively.}
    \label{fig:hist_block_size}
\end{figure}

To examine the threshold of $32$ in Fig. \ref{fig:hr_size_dependence}, we analyzed the histogram of city block sizes (Fig. \ref{fig:hist_block_size}), revealing that almost all blocks are smaller than $160$ m ($32$ grid points). Note that the size of $80$ m ($16$ grid points) is approximately the 75th percentile value for city blocks. At $80$ m, small blocks can be completely contained in cropped data, but larger blocks are always partially contained. Thus, the cropped size of $16 \times 16$ likely makes it difficult to learn large-scale flow modifications, leading to the degradation in accuracy (Fig. \ref{fig:hr_size_dependence}). In the two-stage SR, as large-scale flow corrections are mostly addressed in the first stage, U-Net2 in the second stage only needs to reconstruct street-scale flows and can learn this reconstruction process from cropped small-size data.

\begin{figure}[H]
    \centering
    \includegraphics[width=9cm]{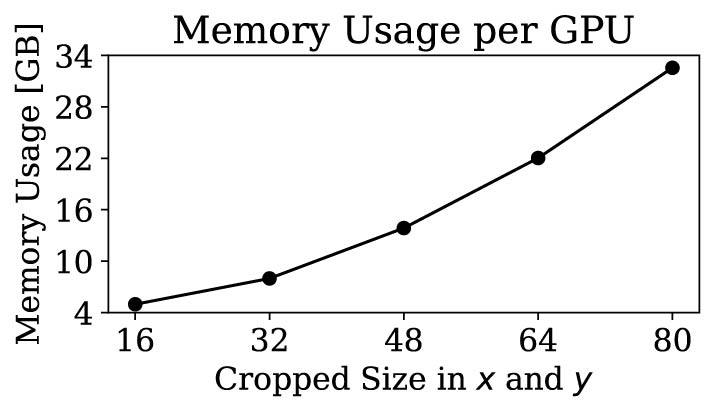}
    \caption{Dependence of GPU memory usage on cropped data size during U-Net2 training. Two NVIDIA A100 GPU boards were used with distributed data parallel. The figure shows memory usage per GPU.}
    \label{fig:gpu_memory_usage}
\end{figure}

This cropping can reduce the necessary GPU memory size for HR data. Figure \ref{fig:gpu_memory_usage} shows GPU memory usage during the second-stage training. The figure indicates that $34$ GB are required for size $80$, while only $4$ GB are needed for size $16$ (about 12\% of $34$ GB). \redA{Clearly, the training over the full domain of size $320$ would exceed the available GPU memory (40 GB). Thus, the two-stage approach would be inevitable if full-domain training was required.} Generally, the necessary spatial size for large-scale flow modification may depend on city layouts, such as block-size statistics, suggesting that the cropped size needs to be carefully determined. In the two-stage SR scheme, however, cropped size can be determined solely by GPU specifications without considering city layouts because large-scale pattern information is not necessary in the second stage.

\section{Conclusions} \label{sec:conclusions}

This study has proposed a two-stage SR simulation method suitable for 3D fluid simulations in urban environments. In this scheme, an NN corrects large-scale flows above buildings in the input LR simulation results; subsequently, another NN reconstructs small-scale flows between buildings \redA{in the HR output}. The proposed method infers street-scale flows without HR numerical integration, thereby significantly reducing computational costs. Due to the scale separation in each stage, the second-stage inference, namely the restoration of HR flows, is spatially localized, which enables training with cropped data of a smaller spatial size. This cropping considerably reduces GPU memory requirements in deep learning.

The proposed method was evaluated using building-resolving micrometeorological simulations for an actual urban area around Tokyo Station in Japan. The HR and LR simulations were conducted under the same initial and boundary conditions, without referencing each other during numerical integration. The two-stage SR successfully inferred the HR flow fields from the input LR simulation results. Compared to the input, the error was reduced by 50\% for air temperature and 60\% for wind velocity. We also verified that the two-stage approach not only localizes SR processes but also stabilizes inference, compared to conventional single-stage SR. The total wall-clock time was reduced to 3.32\% of the HR simulation time (a 30-fold speedup), specifically from 206 min to 6.83 min. These results indicate potential for real-time micrometeorology prediction in urban areas.

\redA{The major limitations of the current study concern temporal and spatial generalizability. Regarding the temporal generalizability, we have focused only on hot days (Section \ref{subsec:micrometeorology}); thus, prediction accuracy in other seasons remains unclear.} Previous SR studies demonstrate that SR simulations provide accurate inferences throughout the year if training data contain atmospheric flows at various times \citep{Oyama+2023SR, Sekiyama+2023AIES}. \redA{This} suggests that the proposed two-stage SR simulation would maintain accuracy \redA{throughout the year if the training data covered all seasons and time periods. Regarding the spatial generalizability, we have examined the SR to the 5-m resolution because most streets are resolved at this resolution. The sensitivity to grid resolution and the SR to higher resolutions should be investigated in future work.}

The two-stage SR will be effective for the future integration of data assimilation, which is necessary for real-time prediction \redA{\citep{Carrassi+2018WIRE}}. Recent significant progress in weather forecasts using NNs is attributable to the availability of assimilated data \citep{Lam+2023Science, Bi2023Nature}. These NNs were trained using assimilated data as a target, which is more accurate than forecast data due to the incorporation of observations. This training method is one reason why NNs can surpass the accuracy of current physics-based meteorology models. Unlike weather forecasts, which target atmospheric flows on kilometer or larger scales, urban micrometeorology involves flows on meter scales and requires substantial computational resources. This computational burden is one reason for the current absence of assimilated data in urban areas. Recent studies have proposed combining data assimilation with SR to improve accuracy \citep{Barthelemy+2022OD,Yasuda+2023JAMES}. Further incorporation of the present two-stage SR will not only improve accuracy but also reduce computational costs. This future integration could develop assimilated data in cities and enable real-time predictions that incorporate observations through data assimilation.

\section*{Declaration of competing interest}

The authors declare that they have no known competing financial interests or personal relationships that could have appeared to influence the work reported in this paper.

\section*{Data availability}

The source code for deep learning is preserved at the Zenodo repository (\url{https://doi.org/10.5281/zenodo.10903319}) and developed openly at the GitHub repository (\url{https://github.com/YukiYasuda2718/two-stage-sr-micrometeorology}). The data that support the findings of this study are available from the corresponding author upon reasonable request.

\section*{Acknowledgements}

The micrometeorological simulation and deep learning were performed on the Earth Simulator system (project IDs: 1-23007 and 1-24009) at the Japan Agency for Marine-Earth Science and Technology (JAMSTEC). This paper is based on results obtained from project JPNP22002, commissioned by the New Energy and Industrial Technology Development Organization (NEDO).

\section*{Declaration of Generative AI and AI-assisted technologies in the writing process}

During the preparation of this work, the authors used Claude 3.5 only for English editing. After using this service, the authors reviewed and edited the content as needed. The authors take full responsibility for the content of the publication.

\appendix

\section{\redA{Hyperparameters of the CNNs}}
\label{sec:hyper-param-appendix}

\redA{We outline the main hyperparameters of the CNNs (U-Net1 and U-Net2). These parameters were determined based on \cite{Yasuda+2023BAE}. All hyperparameter values are available in our full implementation (see Data availability).}

\redA{In U-Net1, the number of channels in convolution is increased through three downsampling blocks as 128, 256, and 512, and is then decreased through three upsampling blocks as 512, 256, and 128 (Fig. \ref{fig:network-architectures}). Similarly, in U-Net2, the number of channels is increased through four downsampling blocks as 128, 128, 256, and 256, and is then decreased through four corresponding upsampling blocks. For both U-Net1 and U-Net2, the convolution kernel size is fixed at 3. The slope of all leaky ReLU functions is $-0.01$ \citep{Maas+2013ProcICML}. The weight parameters in all convolution layers are randomly initialized using uniform distributions \citep{He+2015ICCV}. For the Adam optimizer, we use the default parameters \citep{Kingma+2015ICLR}, except for the learning rates as specified in Section \ref{subsec:training}.}

\redA{We also summarize the data sizes, as they can be considered non-trainable parameters (i.e., hyperparameters), although previously described in Section \ref{subsec:data-preparation}. Each HR (LR) snapshot is represented as a 3D numerical array of size $320 \times 320 \times 40$ ($80 \times 80 \times 10$) along the east, north, and vertical directions, respectively. The input and output physical quantities are listed in Table \ref{table:input-output}. The total number of LR and HR data pairs is 3,420, which is divided into training, validation, and test datasets in chronological order.}

\bibliographystyle{elsarticle-harv}
\bibliography{cas-refs}





\end{document}